\documentclass[prl,twocolumn,showpacs,preprintnumbers,amsmath,amssymb,nofootinbib]{revtex4-1}

\usepackage{graphicx}
\usepackage{dcolumn}
\usepackage{xspace}
\usepackage[utf8]{inputenc}
\usepackage{float}
\usepackage{subfigure}
\usepackage{url}

\usepackage{color}

\newcommand{\eq}[1]{Eq.~(\ref{#1})}

\newcommand{\real}{\mathrm{Re}\,}
\newcommand{\imag}{\mathrm{Im}\,}
\newcommand{\gev}{\,\mbox{GeV}}
\newcommand{\tev}{\,\mbox{TeV}}
\newcommand{\cp}{\mbox{CP}\,}
\newcommand{\eps}{\epsilon}

\newcommand{\no}{\nonumber}
\newcommand{\nn}{\no\\}

\newcommand{\ccb}{c_\beta}
\newcommand{\ssb}{s_\beta}
\newcommand{\ttb}{t_\beta}

\definecolor{BlueViolet}{rgb}{0.2, 0.00, 0.7}
\definecolor{Blue}{rgb}{0.15, 0.00, 0.9}
\definecolor{halayaube}{rgb}{0.4, 0.22, 0.33}
\definecolor{sanddune}{rgb}{0.59, 0.44, 0.09}
\usepackage[colorlinks=true,linkcolor=cyan,citecolor=blue,urlcolor=BlueViolet,hyperfootnotes=false]{hyperref}

\begin{document}

\preprint{TTP--19--50}
\preprint{P3H--19--60}
\title{\boldmath Cornering Spontaneous CP Violation with Charged-Higgs
 Searches}

\author{Ulrich Nierste, Mustafa Tabet, and Robert Ziegler}
\email{ulrich.nierste@kit.edu, mustafa.tabet@kit.edu, robert.ziegler@kit.edu}
\affiliation{\normalsize \it 
 Institut f\"ur Theoretische Teilchenphysik (TTP),
  Karlsruher Institut f\"ur Technologie (KIT), 76131 Karlsruhe, Germany.}

\begin{abstract}
 Decades of precision measurements have firmly established the
 Kobayashi-Maskawa phase as the dominant source of the
 charge-parity (CP) violation observed in weak quark decays.  However,
 it is still unclear whether CP violation is explicitly encoded in
 complex Yukawa matrices or instead stems from spontaneous symmetry
 breaking with underlying CP-conserving Yukawa and Higgs sectors. Here we study the latter possibility  for the case of a generic two-Higgs-doublet model (2HDM). We find
 that theoretical constraints limit the ratio $t_\beta$ of the vacuum
 expectation values (vevs) to the range
 $0.22 \leq t_\beta \leq 4.5$ and imply the upper bounds
 $M_{H^\pm}\leq 435 \gev$, $M_{H_{2}^0} \leq 485 \gev$ and
 $M_{H_{3}^0} \leq 545 \gev$ for the charged and extra neutral Higgs
 masses. We derive lower bounds on charged-Higgs couplings to bottom
   quarks which provide a strong motivation to study the non-standard
   production and decay signatures
   $p p \to qb H^\pm(\to q^\prime b)$ with \textit{all}\ flavors
   $q,q^\prime=u,c,t$ in the search for the charged Higgs boson.  We
further present a few benchmark scenarios with interesting
 discovery potential in collider analyses. 
\end{abstract}
\maketitle
\renewcommand{\thefootnote}{\#\arabic{footnote}}
\setcounter{footnote}{0}
\section{\label{sec:intro}Introduction}
In 1964 the observation of the decay $K_L\to \pi\pi$ has established the
violation of charge-parity (CP) symmetry~\cite{Christenson:1964fg}.
Owing to the CPT theorem~\cite{cpt} this discovery implies that also
time-reversal symmetry (T) is broken and Nature has a microscopic arrow
of time. In 1973 two landmark papers have {proposed} possible mechanisms
of CP violation (CPV) involving new particles: M.~Kobayashi and
T.~Maskawa (KM) pointed out that explicit CPV can occur if the Standard Model
(SM) is amended by a third quark generation~\cite{Kobayashi:1973fv},
while T.D.~Lee showed that spontaneous CPV can be realized in the presence of
  a second Higgs doublet~\cite{Lee:1973iz}.

The subsequent success of the KM mechanism,
however, did not rule out the possibility of spontaneous CP violation: The
complex phase of the Cabibbo-Kobayashi-Maskawa (CKM) matrix 
stems from the diagonalization of complex quark mass matrices, and 
these matrices may still arise as linear combinations of real Yukawa matrices multiplied by complex vevs.    

Almost half a century later, the issue of explicit vs.\ spontaneous CPV
still remains unresolved! The main purpose of this paper is to tackle
this question systematically and {discuss} how to either discover
spontaneous CPV or to entirely rule out this possibility using future
data from precision observables and colliders. The latter is possible,
because spontaneous-CPV scenarios have no decoupling limit {and feature
  a pattern of flavor violation that cannot be aligned to the SM.}

The main obstacle to this endeavor is the considerable size of the
parameter space of SCPV models. Indeed previous works have so far
considered only special cases of 2HDM (see e.g.\
Refs.~\cite{previous,Nebot:2018nqn}).  Our paper targets generic
features of SCPV and only makes two simplifying assumptions, which are
justified by shortcutting to that region of the parameter space that is
least constrained by experiment: Firstly, we identify the lightest
neutral Higgs boson with the 125\gev\ SM-like Higgs particle.  Secondly,
we do not permit Yukawa terms leading to FCNC {Higgs} couplings among
down-type quarks, which are severely constrained by precision flavor
data.  {These data constrain the mentioned couplings so tightly that
  relaxing our second assumption will not change our conclusions.}
We find a remarkable sum rule for charged-Higgs
couplings to $b$-quarks, which implies that at least one of the
couplings to $tb$, $cb$ or $ub$ is sizable. Given the upper limit on the
charged Higgs mass and the constraints from precision observables, these
results reveal that charged Higgs searches in non-standard channels have
the potential to either support or falsify SCPV as the primary origin of the KM
phase.

\section{General features\label{sec:gf}}
\subsection{Higgs sector \label{sec:his}}
The most general  potential with two SU(2) Higgs doublet 
fields $\phi_{i}=(\phi_{i}^0, \phi_{i}^+)^T$, $i=1,2$, reads \cite{Lee:1973iz} 
\begin{align}
& V= m_{11}^2\phi_1^\dagger\phi_1
+m_{22}^2\phi_2^\dagger\phi_2
-(m_{12}^2\phi_1^\dagger\phi_2 + \text{h.c.})  \label{eq:v} \\
&\,\,  + 
\frac{\lambda_1}{2}(\phi_1^\dagger\phi_1)^2 
 +\frac{\lambda_2}{2}(\phi_2^\dagger\phi_2)^2
\nonumber \\ 
 &\,\,
+\lambda_3(\phi_1^\dagger\phi_1)(\phi_2^\dagger\phi_2)
+\lambda_4(\phi_1^\dagger\phi_2)(\phi_2^\dagger\phi_1) 
 \nonumber \\ 
 &\,\,+ \left[  \phi_1^\dagger\phi_2 \left(\frac{\lambda_5}{2}\phi_1^\dagger\phi_2
+\lambda_6(\phi_1^\dagger\phi_1)
+\lambda_7(\phi_2^\dagger\phi_2) \right)
+\text{h.c.}\right], \no
\end{align}
Adopting canonical CP transformation rules, $\cp \phi_{i}(x^\mu)=
\phi_{i}^*(x_\mu) $, CP conservation means that all parameters in
\eq{eq:v} are real. {For appropriate choices of these parameters $V$ is bounded from below and has a local minimum for the complex vevs:}  
\begin{align}
\langle{\phi_1} \rangle & =
\begin{pmatrix}
0 \\ v \ccb
\end{pmatrix} \, , & 
 \langle{\phi_2} \rangle & =
\begin{pmatrix}
0 \\ v \ssb \text{e}^{i\xi}
\end{pmatrix} \, ,
\label{eq:vev}
\end{align}
with $\ccb \equiv \cos \beta >0$,  $\ssb \equiv \sin \beta >0$, $v =174\, \gev$, and the CP phase $\xi$. This minimum is in fact always the global one, which immediately follows from the results in Ref.~\cite{Ivanov:2007de}. As an important observation, the
three minimization equations with respect to $\real \phi_1^0$,
$\real \phi_2^0$, and $\imag \phi_2^0$ allow to trade all three massive
parameters $m_{1}^2$, $m_{2}^2$, $m_{12}^2$ in \eq{eq:v} for the three vev
parameters $v$, $t_\beta \equiv \tan \beta$ and $\xi$. Therefore all elements of the Higgs mass
matrices are of the order of the electroweak scale $v$, with
dimensionless coefficients composed of $\lambda_{1-7}$,
$c_\xi \equiv \cos \xi$, $s_\xi \equiv \sin \xi$ and $t_\beta$. Since
perturbativity does not permit arbitrarily large couplings, the masses
of all Higgs bosons are bounded from above. {This
absence of a decoupling limit in the 2HDM with SCPV has been observed already  
  in Refs.~\cite{Nebot:2018nqn, Nebot:2019lzf}, for other examples see Refs.~\cite{Barenboim:2001vu, Gildener:1976ih,Lane:2018ycs}.}

The Higgs spectrum consists of a charged Higgs with mass 
\begin{align}
m_{H^\pm} =v \sqrt{\lambda_5 -\lambda_4} \, ,  
\end{align}
and three neutral Higgs states $H_A$ with masses that fulfill the sum rule
\begin{align}
\frac{1}{2} \sum_{A=1}^3 \frac{m_{H_A}^2}{ v^2} =s_{2\beta} c_\xi  \left( \lambda_6 +\lambda_7 \right)+\lambda _2 \ssb^2 +\lambda _1 \ccb^2 +\lambda_5 \, .
\end{align}
Requiring NLO perturbative unitarity~\cite{Murphy:2017ojk}
allows to derive upper bounds for the physical Higgs masses, which can
be further tightened by identifying the lightest Higgs with the SM
Higgs.\footnote{We actually require the lightest Higgs
  ${H_1}$ to be  in the mass range $(125 \pm 5) \, {\rm GeV}$. 
  Allowing for new Higgses lighter than the SM
  Higgs weakens the bounds only slightly.}
Using the results in Refs.~\cite{Branco:2011iw,
    Ginzburg:2005dt,lambdabounds}, we find 
\begin{align}
  m_{H^\pm} &\lesssim 435 \gev \, , 
    \label{eq:hpmass}
\end{align}
while neutral Higgs masses must satisfy 
 \begin{align}
m_{H_2} &\lesssim 485\gev \, , &
m_{H_3} &\lesssim 545\gev \, , 
\label{eq:mHbound}
\end{align}
with the sum of all three neutral Higgs masses bounded by $1.1
\tev$. {Our bounds are tighter than those in
  Ref.~\cite{Nebot:2018nqn}, because we include the NLO corrections 
of Ref.~\cite{Murphy:2017ojk}.}
Moreover, since the determinant of the neutral Higgs mass matrix
is proportional to $s_\xi^2 s_{2 \beta}^2$, requiring that all states
are heavier than $125\gev$ gives lower bounds on $s_\xi$ and a range for
$t_\beta$. Using again NLO perturbative unitarity, we find (see
Fig.~\ref{fig})
\begin{align}
0.22 & \lesssim t_\beta \lesssim 4.5 \, , & |s_\xi| & \gtrsim 0.42  \, .
\label{eq:tblimit}
\end{align}
The {neutral} Higgs mass basis is obtained by {diagonalizing}
$O^T M_H^2 O = M_{H,{\rm diag}}^2$ with the orthogonal matrix 
\begin{align}
O \equiv R_{12} (\theta_{12} - \beta ) R_{13} (\theta_{13}) R_{23} (\theta_{23}) \, ,  
\end{align}  
where $R_{ij} (\theta)$ are rotation matrices in the $i-j$ plane by an
angle $\theta_{ij}$. Since the Higgs mass matrices only depend on
$\lambda_{1-7}$ (besides $s_\xi$, $t_\beta$ and $v$), we can trade the
seven $\lambda_i$ parameters for the four Higgs masses $m_{H^\pm}$,
$m_{H_i}$ and the three mixing angles $s_{ij} \equiv \sin
\theta_{ij}$. These mixing angles appear in all couplings of the neutral
Higgs mass eigenstates. The couplings to massive gauge bosons
$g_{H_A VV}$ are given in terms of the corresponding SM Higgs couplings
$g_{hVV}$ by
\begin{align}
g_{H_A VV} = (\ccb O_{1A} + \ssb O_{2A}) g_{hVV} \, .
\end{align}
Particularly simple are the couplings of the lightest neutral Higgs
$g_{H_1 VV}/ g_{hVV} = c_{12} c_{13} $. Since throughout this paper we
will assume that $H_1$ is the observed SM-like Higgs state with a mass
of 125 GeV, its couplings need to be sufficiently close to the couplings
of the SM Higgs, i.e.\ $s_{12}, s_{13} \ll 1 $.


\subsection{Yukawa sector \label{sec:yus}}
The quark Yukawa Lagrangian is given by
\begin{align}
{\cal L}_{\rm yuk}  = - & \overline{Q}_L \left( Y_{u1} \tilde{\phi}_1 + Y_{u2} \tilde{\phi}_2 \right) u_R \nonumber \\
    - & \overline{Q}_L \left( Y_{d1} \phi_1 + Y_{d2} \phi_2 \right) d_R + {\rm h.c.} \, ,
\end{align}
with the Yukawa matrices $Y_{qi}$ and
$\tilde{\phi}_i=\eps_{ij} \phi^*_j, \eps_{12} =1$. Since ${\cal L}_{\rm yuk} $ conserves CP, we can choose $ Y_{q1,q2}$ real.
This implies that fermion mass matrices, given by
\begin{align}
\frac{M_u}{v} & = Y_{u1} \ccb   + Y_{u2}  e^{- i \xi} \ssb  \, , & \frac{M_d}{v} & = Y_{d1} \ccb   + Y_{d2}  e^{ i \xi} \ssb  \, , 
\label{eq:mass_terms}
\end{align}
can induce the KM phase only if $\xi$ is physical, i.e.\ cannot be
rotated away by field redefinitions. This implies flavor misalignment,
defined through $Y_{q1}Y_{q2}^T - Y_{q2}Y_{q1}^T \neq 0$, which
necessarily induces FCNC couplings of neutral Higgs bosons. Since
Eq.~\eqref{eq:mHbound} forbids arbitrarily heavy neutral Higgs bosons,
one cannot suppress all Higgs-mediated FCNC processes simultaneously to
arbitrarily small values. As constraints on FCNC Higgs couplings to
  down-type quarks are particularly strong, in the following we set
  {$Y_{d2} \approx 0$}, thus relegating all FCNC couplings to the up
  sector.  {We stress that this choice is dictated solely by
  phenomenological constraints, and note that it is radiatively stable,
  since loops corrections
  $\delta Y_{d2}\propto Y_{u1} Y_{u2}^T Y_{d1}/(16\pi^2)$ are
  numerically irrelevant.}

Without loss of generality, we can work in a flavor basis where $Y_{d1}$
is diagonal and $M_{u} = V^\dagger m_u^{\rm diag} V_{R}^\dagger$, where
$V$ is the CKM matrix and $V_R$ a free unitary matrix.  The Higgs
couplings to fermions in the mass basis are then given by
\begin{align}\label{eq:LH}
{\cal L}_{\rm H} = &
    - \overline{u}_{L,i} \frac{H^0_A}{\sqrt2} \left[ \delta_{ij} \alpha_A^u + \ttb \tilde{\eps}^u_{ij} \beta_A^u \right]\frac{m_{u_j} }{v \ssb} u_{R,j} \notag \\
    &- \overline{d}_{L,i} \frac{ \alpha_A^d H^0_A}{\sqrt2} \frac{m_{d_i} }{v \ccb}  d_{R,i} \notag \\
    &+ \overline{d}_{L,i} H^-  V_{ki}^* \left[ \delta_{kj} \ccb - \frac{\tilde{\eps}^u_{kj} }{\ccb} \right] \frac{m_{u_j}}{v \ssb} u_{R,j} \notag \\
    &+ \overline{u}_{L,i} H^+ V_{ij} \ssb \frac{ m_{d_j}}{v \ccb}   d_{R,j} + {\rm h.c.} \, , 
\end{align}
with
\begin{align}
\label{eq:alphabetaeps}
  \alpha_A^u & = O_{2A} - i \ccb O_{3A} \,,
& \beta_A^u  & = O_{1A} - \frac{O_{2A}}{t_\beta } + i \frac{O_{3A}}{\ssb} \,, \notag \\
  \alpha_A^d & = O_{1A} - i \ssb O_{3A} \,,
&  \tilde{\eps}^u_{ij} & = (V Y_{u1} V_{R}) _{ij} \frac{v\ccb}{m_{u_j}} \, .
\end{align}
Using Eq.~\eqref{eq:mass_terms}, we can write $Y_{u1}$ as
\begin{align}\label{eq:Yu1}
Y_{u1} &= \frac{1}{v\ccb} \left(\text{Re} + \frac{c_\xi}{s_\xi} \text{Im} \right) \left[ V^\dagger m_u^{\rm diag} V_{R}^\dagger \right] \, ,
\end{align}
which entails an expression for the
couplings $\tilde{\eps}^u_{ij}$:
\begin{align}
\label{eq:epsfinal}
  \tilde{\eps}^u_{jk} & =  \frac{t_\xi - i}{2 \, t_\xi} \delta_{jk} +  \frac{t_\xi + i}{2 \, t_\xi} \left( V V^T m_u V_R^T V_R m_u^{-1} \right)_{jk} \, .
\end{align}
Note that if we use the residual re-phasing freedom to bring
the CKM matrix to the usual Particle Data Group (PDG)
convention $V_{\rm PDG}$, we have $V \to V_{\rm PDG}$ in Eq.~\eqref{eq:LH}, but $V \to V_{\rm PDG} P $ in Eq.~\eqref{eq:epsfinal} with a free
(diagonal) phase matrix $P$. The Higgs couplings only depend on the
combination $V_R V_R^T$, which in this phase convention is a generic
symmetric unitary matrix with three physical phases.  Apart from the angles and phases in $V_R$, all quark flavor violation in the Higgs sector is entirely
determined by up-quark masses and CKM elements.

Taking the lepton Yukawa sector analogous to the down-quark sector, with only one Higgs doublet coupling to right-handed charged
leptons, one obtains a SM-like phenomenology of charged-lepton
decays. The $H^+ \overline{\nu}_{\tau L} \tau_R$ coupling can neither
vanish nor be much larger than $m_\tau/v$, implied by the $t_\beta $
range in \eq{eq:tblimit}.


\subsection{Charged Higgs Couplings\label{sec:charged Higgs}}
\begin{figure}[t]
	\centering
	 \includegraphics[clip,width=.48\textwidth]{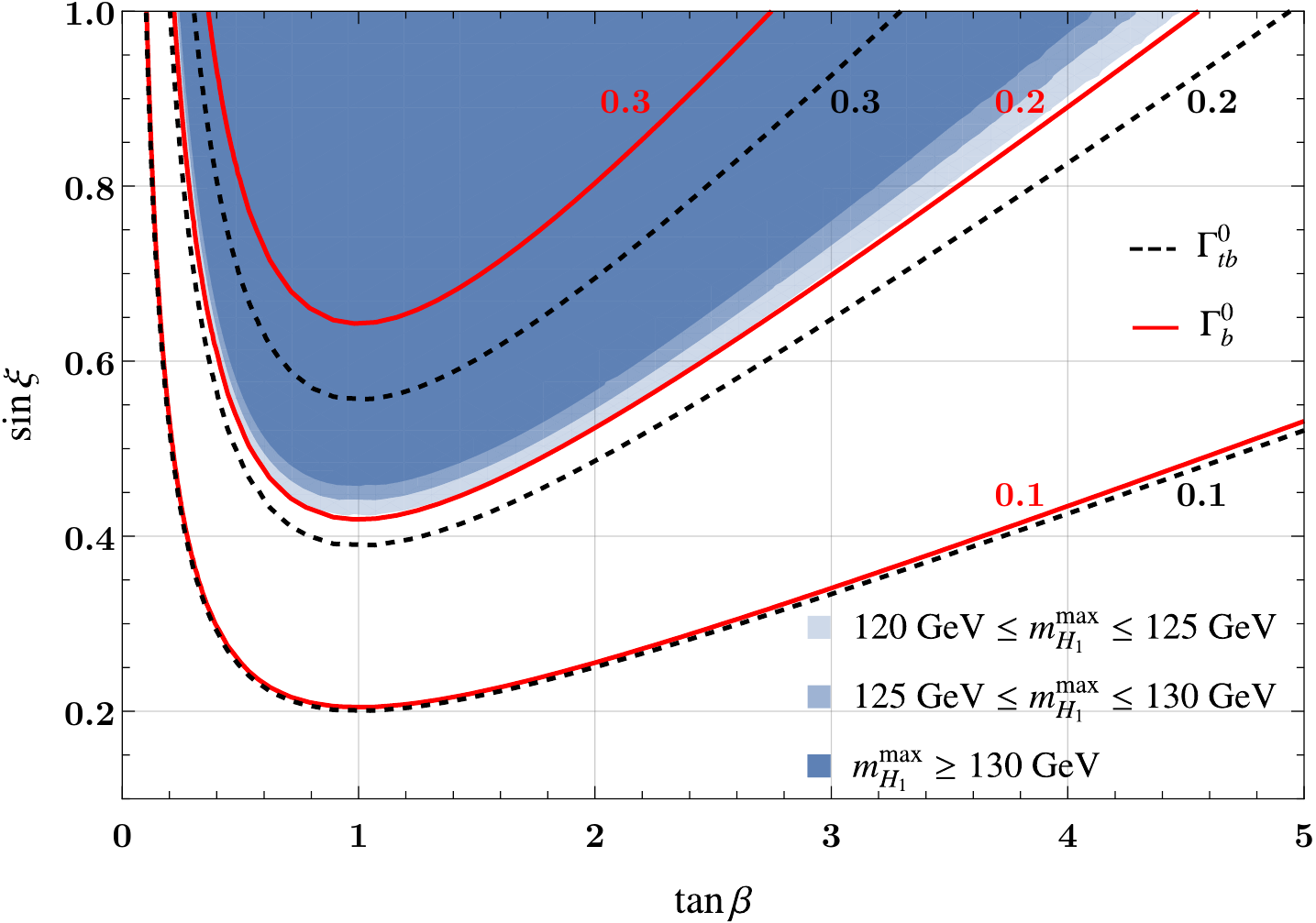}
         \caption{\label{fig} Contours of $\Gamma^0_b$ (red, solid) and
           $\Gamma^0_{tb}$ (black, dashed) in the $t_\beta - s_\xi$
           plane. We also indicate regions with different values of
           $m_{H_1}^{\rm max}$, which is the maximal mass for the
          tree-level value of the lightest Higgs $H_1$ allowed by NLO perturbativity.}
\end{figure}

Since neutral Higgs couplings are more sensitive to the free parameters in $V_R$, we instead focus on the fermion couplings of the charged Higgs. Indeed, the peculiar structure of the Yukawa
sector guarantees that at least one coupling of the charged Higgs to
bottom quarks, $H^+ \overline{u}_{iR} \Gamma^{RL}_{u_i b} b_{L}$, is
sizable. Using Eq.~\eqref{eq:epsfinal} and unitarity of $V_R$, one can show that these couplings satisfy the remarkable relation 
\begin{align}
\label{eq:sumrule}
\!\!\sum_{i= u,c,t} |\Gamma^{RL}_{ib} |^2  \,=\, & \frac{m_t^2}{v^2} +
  \frac{2 m_t}{v s_{2 \beta}} \left( c_{2 \beta}^2 {\rm Re}
  \Gamma_{tb}^{RL} - \frac{{\rm Im} \Gamma_{tb}^{RL}}{t_\xi} \right) \nn
 & \quad + {\cal O}\left( |V_{cb}| \frac{m_c}{m_t} \right)
   \, ,
\end{align} 
{which follows solely from SCPV and the assumption that $Y_{d2}$ is
  approximately diagonal in the same basis as $Y_{d1}$. } This relation
implies that the largest coupling
$\Gamma_{b}^{\rm max} \equiv \max \{ |\Gamma_{ub}^{RL}| ,
|\Gamma_{cb}^{RL}| , |\Gamma_{tb}^{RL}| \}$ is bounded from below
\begin{align}
\label{eq:Gbound}
\Gamma_{b}^{\rm max}  \, \ge\, 
  \Gamma_{b}^{0} \equiv \frac{A}{2 n} \left( \sqrt{1 + n \kappa } -1\right)\, , 
\end{align}
where $n=3$ and the RHS is only a function of $\beta$ and $\xi$
\begin{align}
A & = \frac{2 m_t}{v s_{2 \beta}} \sqrt{ c_{2 \beta}^2 + 1/t_\xi^2 } \, , & 
\kappa = \frac{s_{2 \beta}^2 t_\xi^2}{1 + c_{2 \beta}^2 t_\xi^2} \, .
\end{align}
Minimizing $\Gamma_{b}^{0}$ over $\beta$ and $\xi$ as allowed by NLO
perturbativity and $m_{H_1} \ge 120 \gev$, one numerically finds
\begin{align}
\label{Gbound}
\max \{ |\Gamma_{ub}^{RL}| , |\Gamma_{cb}^{RL}| , |\Gamma_{tb}^{RL}| \} \ge \Gamma^0_b \ge 0.20 \, .
\end{align}
We show contours of $\Gamma^0_b$ in the
$t_\beta - s_\xi$ plane in Fig.~\ref{fig}. As one can see from
this plot, our lower bound on $\Gamma^0_b$ in Eq.~\eqref{Gbound} is
rather conservative.

Note that 
$\Gamma_b^{\rm max}$ reaches its minimum $\Gamma^0_b$ for equal
couplings $|\Gamma_{ib}^{LR}|$, i.e.\ if $\Gamma^{\rm max}_b = |\Gamma_{ib}^{LR}| = \Gamma_{b}^{0}$ for
$i=u,c,t$.  It is instructive to consider two other special cases: If
$\Gamma_{ub}^{LR} = \Gamma_{cb}^{LR} = 0$, then $\Gamma_b^{\rm max}$
coincides with $|\Gamma_{tb}^{LR}|$ and has a minimal value
$\Gamma_{tb}^0$ that is given by the RHS in Eq.~\eqref{eq:Gbound}, but
with $n = 1$. Note that typically $\kappa \ll 1$, which implies that
$\Gamma_{tb}^0$ is only slightly larger than $\Gamma^0_b$. The contours
of $\Gamma_{tb}^0$ in the $t_\beta - s_\xi$ plane are also shown in
Fig.~\ref{fig}, and indeed coincide with those of $\Gamma_{b}^0$ when
$\Gamma_{b}^0$ and therefore $\kappa$ are small. If instead
$|\Gamma_{tb}^{LR}|=0$, the couplings to light generations become large,
since in this case they satisfy the sum rule
$|\Gamma_{ub}^{LR}|^2 + |\Gamma_{cb}^{LR}|^2 = m_t^2/v^2$, which
directly follows from Eq.~\eqref{eq:sumrule}.

The lower bound on charged Higgs couplings to $b$-quarks in
Eq.~\eqref{Gbound}, together with the upper bound on the  charged Higgs
mass in Eq.~\eqref{eq:hpmass} render our class of models predictive
despite the considerable number of free parameters and \eq{Gbound}
entails a ``no-lose'' theorem for charged-Higgs discovery.


\section{Phenomenology\label{sec:pheno}}
{The phenomenology depends on 17 free parameters: 
the heavy Higgs masses $m_{H^\pm}, m_{H_2}, m_{H_3}$, the vacuum angles $\beta$ and $\xi$, the mixing angles
$s_{12}, s_{13}, s_{23}$ and three angles
plus six phases that determine $\tilde{\eps}_u$ in Eq.~\eqref{eq:epsfinal}.  Although huge, this parameter
space is compact because of the absence of new mass scales and
perturbative unitarity, cf.\ Eqs.~\eqref{eq:hpmass}-\eqref{eq:tblimit},
which allows to confirm or rule out the model in the near future. In the following we discuss indirect searches via precision
measurements and direct searches for the new additional Higgs states.}
 
The SM-like measurements of Higgs coupling strengths~\cite{Aad:2019mbh,
  Sirunyan:2018koj} imply small values of $s_{13}$ and $s_{12}$, i.e.\ a
Higgs sector close to the alignment limit. Also constraints from
precision observables like neutral meson mixing~\cite{UTFit,
  Bona:2006sa, Bona:2007vi},
$B\rightarrow X_s\gamma$~\cite{Belle:2016ufb}, and electric dipole
moments (EDMs)~\cite{Afach:2015sja, Andreev:2018ayy} have considerable
impact on the parameter space, {but do not exclude all of it.  Indeed in
  certain, non-trivial parameter ranges all heavy Higgs couplings to
  fermions can be simultaneously suppressed to a level that all
  observables are SM-like. Still many precision observables can be close
  to their current experimental limits, for example electron and neutron
  EDMs can be as large as $|d_e| = 10^{-29} e \, {\rm cm}$ and
  $|d_n| = 3 \cdot 10^{-26} e \, {\rm cm}$, respectively. These regions
  will be explored by several near-future experiments, like
  nEDM~\cite{Ito:2017ywc, Ahmed:2019dhe}, n2EDM~\cite{Abel:2018yeo} and
  the eEDM experiment by the ACME
  collaboration~\cite{Andreev:2018ayy}. Thus precision measurements will
  continue to probe the parameter space from below, pushing up the
  limits on heavy Higgs masses towards the unitarity limits in
  Eqs.~\eqref{eq:hpmass} and \eqref{eq:mHbound}.}

  Also present experimental data from direct Higgs searches constrain
  significant portions of the parameter space, but do not allow to
  exclude the entire scenario. Actually it is quite easy to evade
  standard searches while predicting sizable production cross sections
  for signatures that have not been looked for so far, in particular
  those that result from the dominance of flavor-violating Higgs
  couplings. Indeed it follows from the bound in Eq.~\eqref{Gbound} that
  the charged Higgs is guaranteed to have sizable couplings to bottom
  and up-, charm- or top-quarks. As charged Higgs couplings to gauge
  bosons are suppressed in the alignment limit, while couplings to
  leptons are bounded by the smallness of $t_\beta$, the quark couplings
  typically dominate both production and decay. In the following we
  briefly discuss the resulting charged Higgs phenomenology at the LHC,
  using the benchmark points in Table~\ref{tab:BMtable} as an
  illustration. A much more detailed analysis of the 
collider phenomenology will be presented elsewhere. Because of the upper limit in Eq.~\eqref{eq:hpmass}, we
  are only interested in the light mass range below 440 GeV, which is
  typically more difficult to probe at hadron colliders due to large SM
  backgrounds.

The case of $tb$ associated production and decay to $tb$,
$pp \to tb H^\pm (\to tb)$, belongs to the standard charged Higgs
searches by CMS and ATLAS, cf.\ Refs.~\cite{Khachatryan:2015qxa, Aad:2015typ}
and \cite{Sirunyan:2019arl,Aaboud:2018cwk} for 8 TeV and 13 TeV data,
respectively. 
These searches
exclude signal strengths of ${\cal O}$(1 pb) in the relevant mass range. An
exemplary benchmark point close to exclusion is provided by
\hyperlink{BM1}{BP1} in Table~\ref{tab:BMtable}. Charged Higgs couplings
to $tb$ can be suppressed if couplings to $cb$ or/and $ub$ are enhanced,
which corresponds to fairly unexplored signatures. The phenomenology of
the case of $cb$-dominance is extensively discussed in Ref.~\cite{Gori:2017tvg} (see also
Ref.~\cite{Altmannshofer:2016zrn}). Apart from larger production cross
sections and possible charm tagging in charged Higgs
decays, the case of $ub$ is quite similar to the one of $cb$, so we will
focus on these cases in the remaining discussion, largely following
Ref.~\cite{Gori:2017tvg}. A benchmark point with large $ub$ coupling is
provided in Table~\ref{tab:BMtable} by \hyperlink{BM2}{BP2}.

Starting with $pp \to cb H^\pm (\to cb)$, this process can be probed at
the LHC by inclusive searches for low-mass dijet resonances like
Ref.~\cite{Aaboud:2018fzt}, which however are not yet sensitive to
charged Higgs masses below 450 GeV. Our scenario hopefully motivates
further efforts to optimize future searches for resonances in multi-jet
final states going towards lower masses. For example, we find benchmarks
with (inclusive) production cross sections of $pp \to b (c) H^\pm$ as
large as ${\cal O}$(nb), which are not excluded by present data, see
\hyperlink{BM3}{BP3}.

The next possibility is $pp \to cb H^\pm (\to tb)$, which is represented
by \hyperlink{BM4}{BP4}. Despite the large production cross sections of
${\cal O}$(10 pb) (for the case of untagged charm jets), experimental
searches are hampered by the fact that the jets from the associated $b$-
and $c$-quarks typically fall outside the trigger range for rapidity and
transverse momentum\footnote{See however Ref.~\cite{Ghosh:2019exx} for a recent study of the discovery potential using associated $b$-jets.}. Thus only searches for $tb$ resonances can be used,
which at present focus on the heavy mass range above 1 TeV (see e.g.\
Ref.~\cite{Sirunyan:2017vkm}), and it is unclear whether further data
and optimization will probe masses as low as 300 GeV.

Occasionally $pp \to tb H^\pm (\to cb)$ can be the main production and
decay for charged Higgs masses that are close to the top threshold, see \hyperlink{BM5}{BP5}. The
signature is the same as in LHC searches for
$tt \to Wb H^\pm (\to cb) b$, which so far have been analyzed only for
charged Higgs masses much below the top threshold, see e.g.\
Ref.~\cite{Sirunyan:2018dvm}. Thus our scenario motivates searches also
for masses as large as 170 GeV, together with the models considered in
Ref.~\cite{Altmannshofer:2016zrn, Gori:2017tvg}. 

Other possible signatures like charged Higgs decays into $W H_2$ depend on the details of heavy neutral
Higgs phenomenology, which is more model-dependent. Nevertheless we provide one
 benchmark point \hyperlink{BM6}{BP6} with dominant
$H^\pm \to W^\pm H_2$ decay, where $H_2$ further decays to $\overline{c} c$ or $\overline{b} b$. 
Finally we note that also charged Higgs pair production via Drell-Yan provides a model-independent production channel that varies
between $2\,{\rm fb}$ and $50\,{\rm fb}$ for the benchmark points in
Table~\ref{tab:BMtable}. 


%
\begin{table*}
     \begin{tabular*}{1\textwidth}{ @{\extracolsep{\fill}}c|cccccc}
        \hline\hline
                                       & \hypertarget{BM1}{BP1}             &  \hypertarget{BM2}{BP2}                          &  \hypertarget{BM3}{BP3}               &  \hypertarget{BM4}{BP4}                 &  \hypertarget{BM5}{BP5}      &  \hypertarget{BM6}{BP6}    \\\hline
        $m_{H_2}$ (GeV)                & 170                                & 335                         & 215              & 320                & 160      & 135    \\
        $m_{H_3}$ (GeV)                & 245                                & 355                         & 245              & 335                & 190      & 335    \\
        $m_{H^\pm}$ (GeV)              & 180                                & 375                         & 170              & 325                & 165      & 350    \\\\
        
        $t_\beta$                      & 0.38                               & 0.84                        & 0.43             & 0.67               & 0.36       & 0.37   \\
        $\xi$                          & 1.95                               & 1.96                        & 1.59             & 1.59               & 2.04       & 1.32   \\
        $|s_{13}| \times 10^2$         & 4.4                                &  1.3                        & 0.12             & 0.081              &  7.5       & 1.2    \\
        $|s_{12}| \times 10^2$         & 2.3                                &  1.6                        & 0.32             & 0.095              &  2.8       & 3.4    \\
        $|s_{23}|$                     & 0.23                               &  1.00                       & 0.18             & 0.99               &  0.42      & 0.21   \\\\

        $|\Gamma^{RLH^\pm}_{tb}|$      & $0.38$                             & 0.58    & $0.16$     & $1.17$         & $0.35$      & $0.32$     \\
        $|\Gamma^{RLH^\pm}_{cb}|$      & $<10^{-3}$                         & 0.76    & $0.76$     & $0.70$         & $<10^{-2}$  & $0.37$     \\
        $|\Gamma^{RLH^\pm}_{ub}|$      & $<10^{-6}$                         & 0.45    & $<10^{-4}$ & $<10^{-5}$     & $<10^{-4}$  & $<10^{-7}$ \\\\
        $\sigma_{\rm prod}^{\rm 14 \, TeV}(pp\rightarrow qb H^\pm)$ (pb)            & $\approx 0$         & 190        & 520    & 36     & $\approx 0$ &  7.9 \\
        $\sigma_{\rm prod}^{\rm 14 \, TeV}(pp\rightarrow tbH^\pm)$ (pb)             &  0.62               & 0.28       & 0.13   & 1.6    &  0.71       &  0.10 \\
             $\sigma_{\text{  Drell-Yan }}^{\rm 14 \, TeV}(pp\rightarrow H^\pm H^\mp)$ (fb)    &  35                 & 2.0        & 44     & 3.7    &  49         &  3.5 \\
        Main decay channel of $H^\pm$  & $tb \ 99 \%$ & $cb \ 58 \%$ &  $cb \ 100\%$   & $tb \ 59\%$ & $cb \ 89 \%$ & $WH_2 (\rightarrow c\overline{c}, b\overline{b}) \ 62\%$ \\
        \hline\hline
    \end{tabular*}
      \caption{\label{tab:BMtable} Table with benchmark points compatible with theoretical constraints and current experimental data. The total production cross section at 14 TeV $\sigma_{\rm prod}^{\rm 14 \, TeV}$ in the respective channel has been obtained with
    MadGraph5\_aMC@NLO~\cite{Alwall:2014hca}  using the 
 2HDM\_NLO model file~\cite{Degrande:2014vpa} with the default run
 cards. 
The cross sections for $pp \to qb H^\pm$ denote the sum of the
  inclusive cross sections for a final state containing an untagged $q =
  u,c$ jet. 
}
\end{table*}

\section{Summary and Conclusions}\label{sec:conc}
We have discussed the generic framework of spontaneous CP
violation in the 2HDM, where the KM phase arises solely from the Higgs potential. This scenario has the remarkable
feature that all mass scales are set by the electroweak scale up to
quartic couplings, so that perturbative unitarity implies
model-independent upper bounds on all heavy Higgs states,
cf.\  Eqs.~\eqref{eq:hpmass} and \eqref{eq:mHbound}. Moreover, the new
scalar states must necessarily have  a particular, non-standard
  pattern of flavor violation
in order to induce a non-vanishing  KM phase.  These features
imply that the fate of electroweak SCPV can in principle be determined with
present and near-future experimental data, despite the huge parameter space. The purpose of this paper is to begin this endeavor,
using the most recent results from precision observables and collider
searches.

We have found restricted ranges for Higgs masses and the vacuum angles,
  cf.\ Eqs.~\eqref{eq:hpmass}-\eqref{eq:tblimit}, and have derived
  a lower bound on charged-Higgs couplings to bottom quarks,
  cf.\ Eq.~\eqref{Gbound}. While the remaining parameter space is still huge, it is compact and will be probed from below by precision experiments
  like EDM measurements and from above by neutral and charged Higgs
  searches at colliders.

  In particular the interplay of lower limits on charged Higgs couplings
  and upper limits on the charged Higgs mass leads to large production
  cross sections and branching ratios in channels that have not been
  explored yet. Our framework thus provides a strong motivation for
  non-standard searches at hadron colliders that feature $cb$ or $ub$
  associated charged Higgs production and/or decay. We have provided
  several relevant benchmark points, cf.\  Table~\ref{tab:BMtable}, which
  hopefully stimulate more detailed collider studies of these
  interesting signals that might play an important role in casting the
  final verdict on the origin of CP violation in weak interactions.

\vspace{6mm} {\it Acknowledgments.}--- We thank J.~Zurita and
D.~Zeppenfeld for useful discussions. M.T. acknowledges the support of
the DFG-funded Research Training Group 1694, ``Elementary particle
physics at highest energy and precision''.  The work of U.N. and M.T. is
supported by BMBF under grant no.\  05H2018 (ErUM-FSP T09) - BELLE II:
``Theoretische Studien zur Flavourphysik'' and project C3b of the
DFG-funded Collaborative Research Center TRR 257, ``Particle Physics
Phenomenology after the Higgs Discovery''. {U.N.\ and M.T.\ acknowledge
  the hospitality of the CERN Theory Division, where part of the work
  was done.}

\end{document}